\documentclass[prb,superscriptaddress,twocolumn,showpacs]{revtex4}

\usepackage{graphicx}
\usepackage{amsmath}
\usepackage{amsfonts}
\usepackage{bm}
\usepackage{bbm}

\usepackage{psfrag}

\renewcommand{\P}{\ensuremath{\mathcal{P}}}

\newcommand{\Osc}{\Sigma}
\newcommand{\s}{\sigma}
\renewcommand{\a}{\alpha}
\renewcommand{\b}{\beta}
\newcommand{\ep}{\epsilon}

\newcommand{\epsr}[1]{Fig.~\ref{#1}}

\newcommand{\rvec}{\mathbf{r}}
\newcommand{\Rvec}{\ensuremath{\mathbf{R}}}

\newcommand{\nm}{\text{nm }}

\newcommand{\ket}[1]{\ensuremath{|#1\rangle}}
\newcommand{\bra}[1]{\ensuremath{\langle #1|}}

\newcommand{\rotation}[1]{R}

\newcommand{\refl}{IC_{2y}}
\newcommand{\Cz}{C_{3z}}

\newcommand{\co}{\hspace*{-1ex}(color online) }

\newcommand{\matrixz}[4]{  }  

\newcommand{\eqalN}[3]{                               
  \begin{equation}
      \begin{aligned}
        \label{#1}
      #2\;\;#3
      \end{aligned}
  \end{equation}
 }

\newcommand{\eq}[3]{                               
  \begin{equation}
         \label{#1}
      #2\;\;#3
  \end{equation}
 }

\newcommand{\eqa}[3]{                                
  \begin{eqnarray}     
  #2\;\;#3\;\;
  \end{eqnarray}
 }

\newcommand{\bc}{\begin{center}}
\newcommand{\ec}{\end{center}}
\newcommand{\half}{\frac{1}{2}}
\newcommand{\uu}[1]{\underline{\underline{#1}}}

\newcommand{\bquote}{\lq}    
\newcommand{\equote}{\rq \,} 
\newcommand{\quot}[1]{\bquote #1\equote}

\renewcommand{\vec}[1]{\mathbf{#1}}

\begin{document}

\title{Influence of symmetry and Coulomb-correlation effects on the optical properties of nitride quantum
dots}

\author{N. Baer}
\affiliation{Institute for Theoretical Physics, University of Bremen, 28359 Bremen, Germany}
\author{S. Schulz}
\affiliation{Institute for Theoretical Physics, University of Bremen, 28359 Bremen, Germany}
\author{P. Gartner}
\affiliation{Institute for Theoretical Physics, University of Bremen, 28359 Bremen, Germany}
\affiliation{Institute for Materials Physics, P.O. Box MG-7, Bucharest-Magurele, Romania}
\author{S. Schumacher}
\altaffiliation[Present address: ]{College of Optical Sciences, University of Arizona, Tucson, Arizona 85721, USA}
\author{G. Czycholl}
\affiliation{Institute for Theoretical Physics, University of Bremen, 28359 Bremen, Germany}
\author{F. Jahnke}
\affiliation{Institute for Theoretical Physics, University of Bremen, 28359 Bremen, Germany}

\date{\today}


\begin{abstract}

The electronic and optical properties of self-assembled InN/GaN
quantum dots (QDs) are investigated by means of a tight-binding
model combined with configuration interaction calculations.
Tight-binding single particle wave functions are used as a basis for
computing Coulomb and dipole matrix elements. Within this framework,
we analyze multi-exciton emission spectra for two different sizes of
a lens-shaped InN/GaN QD with wurtzite crystal structure. The impact
of the symmetry of the involved electron and hole one-particle
states on the optical spectra is discussed in detail. Furthermore we
show how the characteristic features of the spectra can be
interpreted using a simplified Hamiltonian which provides analytical
results for the interacting multi-exciton complexes. We predict a
vanishing exciton and biexciton ground state emission for small
lens-shaped InN/GaN QDs. For larger systems we report a bright
ground state emission but with drastically reduced oscillator
strengths caused by the quantum confined Stark effect.
\end{abstract}

\maketitle


\section{Introduction}
The great topical interest in semiconductor nanostructures is not
only based on the realization of some of the paradigm of elementary
quantum mechanics \cite{Mic03,Gru02}, but also by the wide range of
possible applications, ranging from new \cite{GeGa99, MiKi00,
SaPe01} and extremely efficient light sources \cite{ReFo03} to
building blocks for quantum information technology \cite{BoEk00,
ZrBe02,
  Ros04}. Further motivation is provided by the possibility to study
the effect of reduced dimensions on carrier transport \cite{KoAu01}
and optical properties \cite{DeGe98, LaMi98, BaSt00}.

The confinement of the carriers  in all three dimensions on a
nanometer scale is achieved, for example, by embedding regions of a
material with a smaller bandgap in a matrix of a wider bandgap
material. A widespread method of creating such quantum dots (QDs) is
the Stranski-Krastanow growth mode \cite{Noe96, Jac03, StHo04}.

In this fast evolving research field, nanostructures based on
conventional group-III nitrides like AlN, GaN, and InN gained more
interest in recent years. Compared to group-III arsenide
semiconductor materials, nitride-based nanostructures have the
advantage that it is possible to span a much larger spectral range,
presently from amber to ultraviolet, by properly alloying together
the three building blocks and thereby engineering the direct band
gap of these materials \cite{VuMe03}.

Nevertheless and despite the intense research on nitride QDs over
the last decade the understanding of this material system is --
compared to other III-V materials -- still in its infancy.

From the theoretical point of view, one challenge is the proper
inclusion of effects that stem from the altered atomistic structure
of the underlying lattice. While most of the nitrides can
crystallize both in the zinc-blende and the wurtzite phase, the
latter is by far more stable \cite{Mor99}. Additionally, the  strong
built-in field needs to be considered for, the mostly applied growth
along the $c$-axis. In contrast to many other III-V semiconductors,
the spin-orbit coupling in the nitrides is rather weak \cite{VuMe03}
so that the calculation of the electronic states in terms of a
simple effective-mass approximation is not possible due to strong
valence band mixing effects. Instead, especially for small
nanostructures, a microscopic description of the single-particle
states based on, for example, a tight-binding model or
pseudo-potential calculation is necessary.

In the past many experimental and theoretical investigations
addressed the optical properties of III-V and II-VI QD structures
based on, e. g., InGaAs/GaAs or CdSe/ZnSe. One important aspect
concerned the analysis of absorption and emission spectra of QDs as
a function of the excitation density~\cite{BaDu95, HaWo96, DeGe98,
DeGe98b, Haw99, BaSt00, HoMo00, BaOr02, BeNa03, Mic03}.  A common
result is that the optical processes mainly involve 'diagonal'
transitions, that are connecting, for example, $s$-shell electrons
with $s$-shell holes or $p$-shell electrons with $p$-shell holes. In
envelope-function approximation this is traced back to dipole matrix
elements calculated from the Bloch functions, and interband
transition amplitudes determined by the product of dipole matrix
elements and the overlap of the electron and hole envelope
functions. This picture has been proven to be very fruitful for
conventional III-V materials and is oftentimes carried over to the
nitride system where emission spectra are then calculated using
these `diagonal'
excitons.~\cite{MaDo99,HoFe99,RiAm02,Shi02,ChSc02,RiAm04} Our
analysis shows, however, that for nitride QDs deviations from this
picture are possible. We identify situations, in which the emission
is dominated by recombination of `skew' excitons, that are excitons
which consist of $s$-shell electrons and $p$-shell holes or vice
versa. This led to the prediction of dark exciton and biexction
ground state emission.~\cite{BaSc05} Such a modified selection rule
cannot be explained in an simple effective-mass picture but requires
the inclusion of band-mixing effects in a way that properly accounts
for the symmetry of the underlying atomic lattice. In addition to
the important changes of the optical properties involving one and
two excitons, the transitions that occur in systems with a higher
population of excitons differ strongly from those known from the
InGaAs/GaAs system. While for smaller nitride QDs the ``skew''
excitons are predicted, we find that in large nitride QDs the
energetic order of the $s$-shell and $p$-shell for holes is
interchanged, with the $p$-shell being lowest in energy. This leads
to optically active exciton transitions.~\cite{ScSc06} However, as a
result of the quantum confined Stark effect (QCSE), with increasing
height of the QDs the corresponding  dipole matrix elements decrease
drastically~\cite{AnRe01, RiAm02,ScSc06} due to the separation of
the electron and hole wave functions in the strong internal electric
field. In combination with the dark exciton state in small nitride
QDs, this makes the application of nitride QDs more challenging.

In this work we study the multi-exciton emission spectra in nitride
QDs and discuss the resulting complicated peak structure in detail.
Restricting ourselves to the two lowest shells allows even a
semi-analytic description of the problem. The single-particle states
are deduced from an atomistic tight-binding model as briefly
outlined in Sec.~\ref{TB_stefan}. Even without a detailed
calculation of the single-particle states, the symmetry
considerations of Sec.~\ref{sym_wurtzite} can be used to draw
conclusions regarding the dipole-matrix elements and the
multi-exciton spectra. In Sec.~\ref{sec:manybodyprob} the
Hamiltonian of the interacting charge carrier system and the
configuration interaction (CI) scheme are presented. Results for an
initial filling with up to six excitons are provided in
Section~\ref{nitride_FCI} and major trends are discussed in
Section~\ref{nitride_diag}  in terms of an approximate Hamiltonian.
Further details of the multi-exciton spectra are analyzed in
Section~\ref{sec:semianalytic} In the subsequent
Section~\ref{field_vs_nofield} the influence of the strong built-in
field  is investigated. Finally, the spectra for a larger nitride QD
are presented in Section~\ref{larger_dot} and compared to those
found for the smaller structure.

\section{Electronic structure}
\label{TB_stefan}

For a proper treatment of the single particle states in nitride
systems, an atomistic multiband approach, like pseudopotential or
tight-binding (TB) models, is required. General aspects for the
developement of such a TB approach and the calculation of dipole and
Coulomb matrix elements are discussed in detail in
Ref.~\onlinecite{ScSc06}. In the following Section~\ref{TBfield}, we
briefly summarize the main ingredients of the subsequently used TB
model, while Sections~\ref{sec:singlestates} and
~\ref{interactionmatrix} are devoted to details of the calculation
of single-particle states and interaction matrix elements.

\subsection{Tight-Binding model and built-in field}
\label{TBfield}

For the calculation of the single-particle states we use a TB model
with an $sp^3$ basis $|\alpha\rangle_\mathbf{R}$, that contains one
$s$-state ($\alpha= s$) and three $p$-states ($\alpha= p_x, p_y,
p_z$) per spin direction at each atomic site $\mathbf{R}$. The
$i^\mathrm{th}$ TB single-particle wave function is given by
\begin{equation*}
|\phi_i\rangle=\sum_{\alpha,\mathbf{R}}|\alpha\rangle_\mathbf{R}\,c^{i}_\alpha(\mathbf{R})
\end{equation*}
with the expansion coefficients denoted by
$c^{i}_\alpha(\mathbf{R})$. We assume a lens-shaped InN QD, grown in
(0001)-direction on top of an InN wetting layer (WL), that is
embedded in a GaN matrix. For the numerical evaluation, a finite
wurtzite lattice within a supercell with fixed boundary conditions
is chosen. The small spin-orbit coupling and crystal-field splitting
are neglected~\cite{VuMe03}. Furthermore, we neglect the strain
induced displacement of the atoms. For the used QD geometry, a more
realistic inclusion of strain effects does not change the symmetry
of the system and henceforth the general statements discussed in
this paper should not be affected. In the following we consider two
different QD sizes with diameters $d=4.5, 5.7$ nm and heights
$h=1.6, 2.3$ nm, respectively. In both cases a WL thickness of one
lattice constant $c$ is assumed.

In contrast to cubic III-V semiconductor heterostructures, e.g.,
InAs or GaAs, the III-V wurtzite nitrides exhibit considerable
larger built-in fields~\cite{Bernardini1997}, which can
significantly modify both the electronic and the optical properties.
This field enters via the electrostatic potential as a site-diagonal
contribution $V_p(\mathbf{R})=-e\phi_p(\mathbf{R})$ to the TB
Hamiltonian.~\cite{Sala1999,Saito2002}

\begin{figure}
\begin{center}
\includegraphics[totalheight=5.25cm]{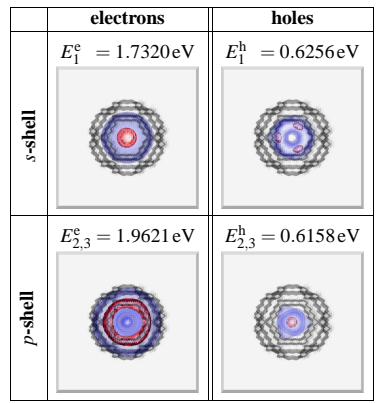}
\caption{\label{oneparticle}(color online). Top view of the QD
structrue with the first two bound shells for electrons (left) and
holes (right). Depicted are isosurfaces of the probability density
with 20\% (blue) and 80\% (red) of the maximum value are shown. Only
one state is visualized for the two-fold degenerate $p$-shell, as
the other looks alike. The corresponding energies
($E^{e,h}_{1,2,3}$) are measured relative to the valence band
maximum of the bulk GaN.}
\end{center}
\end{figure}

\subsection{Single-particle states}
\label{sec:singlestates} The small QD ($d=4.5$ nm, $h=1.6$ nm)
confines three bound states for the electrons. The corresponding
probability densities of the first two bound shells for electrons
and holes are presented in Fig.~\ref{oneparticle}. In the
calculation, the influence of the built-in field was included. The
dominant contribution for the electron single-particle states stems
from the atomic $s$-orbitals, while for the hole states a strong
intermixing of the atomic $p$-orbitals is observed. This is
indicative for the fact that a suitable description of the
one-particle states and the resulting optical properties requires a
multi-band approach and cannot be accounted for in a one-band
effective mass approximation.

If one compares the single-particle states with and without the
inclusion of the built-in field,~\cite{ScSc06} one finds that, in
the presence of the field, the electron states are squeezed into the
cap of the QD while the hole states are constraint to a few atomic
layers at the bottom near the WL. This effect is even more
pronounced for the larger QD ($d=5.7$ nm, $h=2.3$ nm). In this case
a clear spatial separation of electron and hole wave functions is
observed, which lowers the direct spatial overlap between the two
and leads to strongly reduced dipole matrix elements. Furthermore,
as known from the QCSE, the built-in field leads to an overall
red-shift in the single-particle transition energies. In addition,
we find for the large QD, that the ordering of the energy levels is
affected by the built-in field. Specifically we obtain that in the
presence of the built-in field the level ordering for the holes is
reversed so that the ground state is formed by the two-fold
degenerate states $\phi^h_{2,3}$ ($p$-shell in
Fig.~\ref{oneparticle}) while the first excited state $\phi^h_{1}$
(s-shell in Fig.~\ref{oneparticle}) is nondegenerate. This is in
contrast to the observed level ordering of the large QD in the
absence of the built-in field or of the small QD with and without
the field. In these cases, the ground state is nondegenerate
($\phi^h_{1}$) and the first two excited states are degenerate
($\phi^h_{2,3}$). Such a behavior has been reported before for other
QD systems and a detailed discussion can be found in
Ref.~\onlinecite{ScSc06}.

%
%
\subsection{Evaluation of dipole- and interaction matrix elements}
\label{interactionmatrix} Based on the TB single-particle
wave-functions one can determine dipole- and Coulomb matrix elements
that are crucial for the calculation of optical properties. As
emphasized above, a TB model represents an atomistic approach to
describe the electronic structure of low-dimensional
heterostructures. Explicit knowledge about a basis set of localized
(Wannier) states is not required for the calculation of one-particle
energies and wave functions, because the Hamiltonian matrix elements
between the different orbitals are treated as parameters within the
TB model. What enters the TB calculation are only the basic
assumptions about the localized (atomic) orbitals: symmetry, spatial
orientation \cite{Slater1954}, and orthogonality.

For the calculation of interaction matrix elements one needs -- at
least in principle -- the localized basis states. For the Coulomb
matrix elements, however, the explicit knowledge of the atomic
orbitals is not required in practice, as these matrix elements are
dominated by long-range contributions in which the local orbitals
act as point charges. The structure of the localized orbitals is of
significance only for on-site and nearest-neighbor interactions,
which in our calculation contribute less than $5\,\%$ to the total
Coulomb matrix elements. These findings are in agreement with
Ref.~\onlinecite{SuCh00}. Thus, the matrix elements are approximated
by a sum over the TB coefficients $c_{\alpha}^{i}(\mathbf{R})$ at
atom sites $\mathbf{R},\mathbf{R}'$ with orbital indices
$\alpha,\beta$:

\begin{align}
\nonumber
  V_{ij,kl} & = \int d^3 r  \int d^3 r' \;
       \phi_{i}^{*}(\vec r) \phi_{j}^{*}(\vec r^{\,\prime})
       V(\vec r -\vec r^{\,\prime})
       \phi_{k}^{}(\vec r^{\,\prime})\phi_{l}^{}(\vec r)   
\\ \nonumber
& \approx\sum_{\mathbf{R}
\mathbf{R}'}\sum_{\alpha\beta}c_{\alpha}^{i\ast}(\mathbf{R})
c_{\beta}^{j\ast}(\mathbf{R}')c^k_{\beta}(\mathbf{R}')c^l_{\alpha}(\mathbf{R})\,V(\mathbf{R}-\mathbf{R}')
\\\nonumber
\text{with}&\quad V(\mathbf{R}-\mathbf{R}')=
\frac{e^2}{4\pi\varepsilon_0\varepsilon_r|\mathbf{R}-\mathbf{R}'|}\quad\text{for}
\quad \mathbf{R}\not=\mathbf{R}'
\\\nonumber\label{EqCoulappr1}
\text{and}& \quad
V(0)=\frac{1}{v^2_{uc}}\int_{uc}d^3r\,d^3r'\frac{e^2}{4\pi\varepsilon_0\varepsilon_r|\mathbf{r}-\mathbf{r}'|}
\; .
\end{align}

The labels $i,j,k,l$ refer either to electron or to hole states in
order to consider the repulsive electron-electron and hole-hole
interaction, as well as the attractive electron-hole interaction.
The considerably smaller matrix elements of the electron-hole
exchange interaction are neglected. The calculation of the onsite
integrals $V(0)$ involves the integration over the volume of the
unit cell $v_{uc} $ and can be done quasi-analytically by expansion
of the Coulomb interaction in terms of spherical
harmonics.~\cite{ScCz02} The electronic charge and the vacuum
dielectric constant are denoted by $e$ and $\epsilon_0$,
respectively. We use the InN dielectric constant
$\varepsilon_{r}=8.4$ according to Ref.~\onlinecite{Shi2003} since
the wave functions are almost completely confined inside the QD.

For the calculation of dipole matrix elements
$\mathbf{d}^{eh}_{ij}\propto \langle\phi^e_i|\mathbf{r}|\phi^h_j
\rangle$, the explicit structure of the localized orbitals is needed
as the dipole-operator
is dominated by short-range contributions. Though being not
orthogonal at different sites, standard Slater orbitals
\cite{Slater1930} have been used in earlier
calculations~\cite{Lee2001,Lee2002} within orthogonal TB models.
While they include the correct symmetry properties, the missing
orthogonality limits their applicability. To overcome this problem,
we use numerically orthogonalized Slater orbitals~\cite{ScSc06} and
account for the slight non-locality of the dipole
operator~\cite{LePo1998} and the underlying anion-cation structure
of the crystal by including also contributions from up to
second-nearest neighbors.

\section{Symmetry considerations}
\label{sym_wurtzite}
%
%
%
%
The specific symmetry of the system under consideration plays an
important role in the prediction of energy degeneracies and optical
selection rules. The overall symmetry of the problem is determined
by two factors: (i) the crystal symmetry of the underlying lattice
and (ii) the symmetry  of the QD geometry or, more generally, the
geometry of the heterostructure.\cite{BeZu05}

The interplay between the dot geometry and the underlying lattice is
well illustrated by the example of the lens-shaped QD. The
\emph{geometry} of the QD has a $C_{\infty v}$
symmetry.\cite{c_infty} In the effective mass approximation  or in
the $\vec{k}\cdot\vec{p}$ theory the underlying atomic structure is
not resolved, so that one obtains degenerate energies for the
lens-shaped QD.~\cite{WoHa96,Jacak,BeZu05} However, if one takes the
\emph{crystal structure} into account, as it is done, in a
TB\cite{ScCz05,ScSc06} or pseudo-potential calculation
\cite{BeZu03}, the symmetry will be reduced and degeneracies can be
lifted.\cite{BeZu05}
For a lens-shaped QD grown on a wurtzite lattice along the
(0001)-direction, the symmetry is reduced to $C_{3 v}$. If a QD with
the same geometry is grown in the zinc-blende phase in
(001)-direction, one is left with an even lower $C_{2 v}$ symmetry.
As we will discuss in the next subsection, the former group is still
rich enough to predict degeneracies, while the latter is too small
to support degenerate eigenstates.
%
%
%
%

\subsection{Energy spectrum}
\label{nitride_deg}

The $C_{3v}$ group is generated by the rotation around the $z$-axis
with an angle of $2\pi/3$, denoted by $\Cz$, together with a mirror
reflection at a plane perpendicular to the $y$-axis, denoted by
$\refl$.~\cite{Cor69} These operations commute with the Hamiltonian,
but not among themselves. From this non-abelian character of the
group one can immediately conclude that there must exist
energetically degenerate states. In the case of the wurtzite QD
under consideration, one example of such degenerate states  are the
$p$-shell levels given in Fig.~\ref{oneparticle}.

This degeneracy in the wurtzite structure is especially worth
mentioning in view of a recent paper~\cite{BeZu05} in which it is
argued that a description at the atomic level in zinc-blende-based
QDs should remove the $p$-shell degeneracy. The argument is
essentially the following: the four-fold rotation is part of the
zinc-blende point group only if followed by the inversion, but many
high symmetry (lens-shaped, pyramidal) QDs forbid the inversion and
with it the $C_{4 v}$ symmetry. One is therefore left with a $C_{2
v}$ point group, which is abelian and therefore insufficient for
$p$-degeneracy.

This conclusion is correct, but it cannot be carried over to the
wurtzite QDs, because in wurtzite crystals the three-fold rotation
holds without inversion. If the QD geometry is sufficiently
symmetric (lens-shaped, hexagonal pyramids) then the problem retains
the $C_{3 v}$ symmetry, which leads to $p$-degeneracy. In papers
predicting $p$-shell splitting for QDs in wurtzite lattices and
neglecting the spin-orbit coupling~\cite{AnRe00, FoBa03}, either
boundary conditions or discretization meshes spoil the correct $C_{3
v}$ symmetry of the problem. Therefore the small $p$-shell splitting
obtained there is a numerical artefact. On the contrary the
splitting induced by the inclusion of the spin-orbit
coupling~\cite{WiSc06}, albeit quite small, is physically correct.

To summarize the discussion, for a lens-shaped QD and in the absence
of the spin-orbit coupling a degeneracy of the $p$-shell is expected
in the wurtzite phase but not in a zinc-blende structure.

\subsection{Single-particle states}

So far we discussed only the role of symmetry on the
energy spectrum. Turning now to the dipole selection rules, one
should consider the symmetry properties of the wave functions.
In the TB description the wave functions are expanded in the atomic
orbital basis according to
\begin{equation}
\label{eq:TB_wave_symmetry} |\phi\rangle = \sum_{\Rvec}
|s\rangle_{\Rvec} \; a(\Rvec) + \sum_{\Rvec,j} |{p_j}\rangle_{\Rvec}
\; b_j(\Rvec)\,\, ,
\end{equation}
where $|s\rangle_{\Rvec}$ denotes the atomic $s$-orbital centered
around the  site $\Rvec \:$ with $ \langle\rvec|s\rangle_{\Rvec} =
\psi_s(\rvec-\Rvec)$, and the corresponding $p$-orbitals  follow
similarly from $|p_j\rangle_{\Rvec}$ with $j= x,y,z$. The expansion
coefficients are given by $a(\mathbf{R})$ and $b_j(\mathbf{R})$,
respectively.

The action of a symmetry operation $T$ on the wave function is defined
via~\cite{InTa90, Cor69}
\begin{equation}
\label{transf} (T \phi)(\rvec)=\phi(T^{-1} \rvec).
\end{equation}
In the case of the TB wave function,
Eq.~(\ref{eq:TB_wave_symmetry}), this amounts to a simultaneous
transformation of the orbitals and of their centers. Instead of the
original orbitals $p_x$ and $p_y$, it is more convenient to work
with the complex combinations $p_\pm= (p_x\pm ip_y)/\sqrt{2}$ which
are angular-momentum eigenstates.

For a wurtzite QD with $C_{3v}$ symmetry  one has to consider the
transformation properties of the orbitals under the discrete
$2\pi/3$ rotation $\Cz$ around the $z$-axis and the mirror
reflection  $\refl $, defined by $(x,y,z)\rightarrow(x,-y,z)$. While
the atomic $s$ and $p_z$ orbitals do not change under these
transformations, the remaining $p_\pm$ orbitals transform like
\begin{equation}
\begin{aligned}
\Cz \ket{p_+} &= e^{-i\frac{2\pi}{3}}\ket{p_+}\;,
 \quad &  \refl \ket{p_+} &= \ket{p_-}\;, \\
\Cz \ket{p_-} &= e^{+i\frac{2\pi}{3}}\ket{p_-}\;,
 \quad &  \refl \ket{p_-} &= \ket{p_+}.
\end{aligned}
\label{p_pm}
\end{equation}
Using Eq.~(\ref{transf}) these transformation properties of the
\emph{local} orbitals under the group basic rotations are carried
over to the \emph{total} wave  function, given by
Eq.~(\ref{eq:TB_wave_symmetry}). This defines the action of the
group on the TB basis and allows the classification of the
eigenstates according to the irreducible representations of the
$C_{3v}$ group. States which are invariant under the action of the
group will be called  $s$-states and will be denoted by
$\ket{\phi_s}$. This terminology and notation are somewhat lax since
according to it a $p_z$-orbital is also an $s$-state. Nevertheless
we use them for the sake of simplicity and in agreement with the
literature. Similarly, the basis vectors of a two-dimensional
representation changing under the group operations like in
Eq.~(\ref{p_pm}) will be called $p_\pm$-states and will be denoted
$\ket{\phi_{p_\pm}}$.

As an example, it can be shown that an $s$-state wave function must
have the form
\begin{eqnarray}\label{TB_s_state}
\ket{\phi_s} & = &    \sum_{\Rvec}   \big[\;  \ket{s}_{\Rvec}
  \; \alpha(\Rvec)
+ \ket{p_z}_{\Rvec} \;\beta_z(\Rvec)  \; \big] \\
\nonumber
& & +  \sum_{\Rvec} \big[\; \ket{p_+}_{\Rvec}\;
\mathcal{Z}^*_{\Rvec} \;\beta(\Rvec) +  \ket{p_-}_{\Rvec}\;
\mathcal{Z}_{\Rvec}\; \beta^*(\Rvec) \big]
\end{eqnarray}
where the coefficients $\alpha(\Rvec)$, $\beta_z(\Rvec)$, and the
real part of $\beta(\Rvec)$ are invariant under all rotations of
$C_{3v}$, while the imaginary part of $\beta(\Rvec)$ is invariant
under the proper rotation $\Cz$  but changes sign under the action
of the improper rotation $\refl$. The quantity $\mathcal{Z}_{\Rvec}$
is defined as ${X}_{\Rvec} + i {Y}_{\Rvec}$ or in polar coordinates
$\sqrt{ X_{\Rvec}^2 + Y_{\Rvec}^2} e^{i\xi_{\Rvec}}$, where
${X}_{\Rvec}$ and ${Y}_{\Rvec}$ denote the in-plane cartesian
coordinates of the lattice site $\Rvec$ and $\xi_{\Rvec}$ is the
polar angle of the point $\Rvec$ in cylindrical coordinates. It is
clear that with $\alpha(\Rvec)$, $\beta_z(\Rvec)$ rotation invariant
the first sum in Eq.~(\ref{TB_s_state}) is $s$-like. In the second
sum one has a compensation between the $p$-behavior of the orbitals
and that of the $\mathcal{Z}_{\Rvec}$ coefficients leading to a
combination which is left invariant by the rotations. The equivalent
of this situation in a $\vec{k}\cdot\vec{p}$ approach would be
expressed by the $p$-character of the envelopes multiplying to the
$p_{\pm}$-type Bloch functions and compensating the phase factor
acquired  under rotation in order to produce full invariance. In
this way, the coefficients of the expansion Eq.~(\ref{TB_s_state})
can be seen as discretized versions of the envelope function
components.

For the electron and hole wave functions whose modulus square is
depicted in \epsr{oneparticle} the lowest-lying levels are indeed
invariant under the action of the elements of the discrete group and
therefore represent $s$-states. As expected, the electron $s$-state
consists mainly of $s$ atomic orbitals, i. e. the dominant
coefficient is $\alpha(\Rvec)$, which is $C_{3v}$-invariant as
stated above. This state looks very similar to what would be
expected for an $s$-state in the effective mass approximation: it is
symmetric under rotation, has a single maximum at the center and
decays to the boundaries. In contrast, the shape of the ground state
of the holes does not show the same behavior. It shows the
(discrete) rotational symmetry, but has a node at the center. This
state is expected to consist mainly of $p$ atomic orbitals and
indeed, an inspection of its coefficients shows that the second sum
in Eq.~(\ref{TB_s_state}) is dominant. Since ~\epsr{oneparticle}
displays the sum of the modulus square of the expansion
coefficients, one obtains $2|\mathcal{Z}^*_{\Rvec} \;\beta(\Rvec)|^2
= 2  ( X_{\Rvec}^2 + Y_{\Rvec}^2)|\beta(\Rvec)|^2 $, which explains
both the rotational symmetry and the vanishing value at the origin.
In other words the unexpected node at the origin stems from the
$p$-character of the coefficients.

A similar analysis of the coefficients can be carried out for the
excited states ($p$-shell), and reveals that these states have indeed
the $\ket{\phi_{p_\pm}}$ symmetry.

\subsection{Dipole-matrix elements\label{sec:dipolesymm}}

The numerical evaluation of the dipole-matrix  elements shows that
the non-zero values of the in-plane dipole matrix elements
$\vec{e}\vec{d}^{eh}_{ij}$ with $\vec{e}$ in the $x$-$y$-plane are
much larger than those with $\vec{e}$ in $z$-direction. Therefore we
consider here only the in-plane matrix elements with
$\vec{e}=1/\sqrt{2}(1,1,0)$ and denote the corresponding
dipole-matrix elements by $d^{eh}_{ij}$. Furthermore the matrix
elements ${d}^{eh}_{p_ip_j}$, with $i,j\in\{x,y\}$ are more than one
order of magnitude smaller than the other non-vanishing matrix
elements. As a consequence, these contribution can safely be
neglected in the calculation of optical spectra, in which the
absolute value of the dipole-matrix elements enters even
quadratically. As explained below, it turns out that these results
can be understood on symmetry grounds.


To this end we consider the matrix elements $\bra{\l_z} x\pm
iy\ket{\l'_z}$, from which all the in-plane dipole matrix elements
can be deduced by linear combinations and are characterized by
simple transformation properties. Here we denote the states
$|l_z\rangle$ by the phase factor $e^{-i\l_z\theta}$ they pick up
under a rotation around the $z$-axis of the angle $\theta$, that is,
$\l_z=0$ for the $s$-state and  $\l_z=\pm 1$ for the $p_\pm$-states.
Using the transformation properties of the wave functions and those
of $x \pm iy$ under rotation, one can rewrite the matrix elements as
\begin{equation}
\bra{\l_z} x \pm iy\ket{\l'_z} =  e^{ i(\l_z-\l_z'\mp 1) \theta
}\bra{\l_z}   x \pm iy \ket{\l_z'} {.}
\end{equation}
In a $C_{\infty v}$ group the angle $\theta$ is arbitrary and one
would obtain the familiar result that the dipole matrix elements
$d^{eh}_{\l_z,\l_z'}$ can only be non-zero for $\l_z-\l_z'=\pm 1$.
In particular one has $d^{eh}_{ss}=0=d^{eh}_{p_i,p_j}$ with $i,j\in\{\pm 1\}$
and  only $d^{eh}_{s,p_i}$ and $d^{eh}_{p_i,s}$ can be non-zero.

In case of the $C_{3v}$ symmetry the allowed values for the angle
$\theta$ are only the integer multiples of $2\pi/3$. Then  one finds
the weaker result that the condition for non-zero dipole matrix
elements is $\l_z-\l_z' =\pm 1 \text{ modulo }3 $. This means that
the discussed symmetry reduction opens additional decay channels.
One  still has $d^{eh}_{ss}=d^{eh}_{p_+p_+}=d^{eh}_{p_-p_-}=0$ but,
in addition to  $d^{eh}_{p_\pm, s}$, $d^{eh}_{s,p_\pm}$, now also
the matrix elements $d^{eh}_{p_+p_-}$ and $d^{eh}_{p_-p_+}$ can be
non-zero. Nevertheless, since the last two are vanishing in case of
a continuous rotation symmetry, it is quite plausible that they
remain small even in the case of  a three-fold axis. This behavior
is found by the numerical evaluation of the dipole matrix elements.
In our example of the small QD we find
$|d^{eh}_{p_+p_-}|^2=4.84\times 10 ^{-4}$ $(\text{\AA})^{2}$, which
is negligible in comparison with
 $|d^{eh}_{sp_\pm}|^2 = 4.73$ $(\text{\AA})^2$.

As we will see in Section~\ref{nitride_excitonic_properties} these
selection rules, and in particular the vanishing dipole matrix
element $d^{eh}_{ss}$, have very important consequences for the
optical properties of nitride QDs.

\subsection{Coulomb matrix elements}

\begin{table}
\caption{All nonzero electron-electron Coulomb matrix elements
determined numerically from the TB-wave functions. The index 0
denotes the $s$-state $|\phi_s\rangle$ and $\pm$ denote the states
$|\phi_{p_{\pm}}\rangle$. The matrix elements below the horizontal
double line would be zero in the case of a $C_{\infty v}$ symmetry.
Explicit values are given for of the small the small QD but without
the inclusion of the internal electrostatic field.\\}
\begin{tabular}{c c}

\textbf{(i,j,k,l)} &  \textbf{$V^{ee}_{ijkl}$ [meV]} \\
\hline

(0,0,0,0) & 93.8459 \\ \hline


(0,+,+,0) , (0,-,-,0) , (+,0,0,+) , (-,0,0,-) & 81.6389 \\ \hline



(-,-,-,-) , (+,+,+,+) , (+,-,-,+) , (-,+,+,-) & 75.8542 \\ \hline

(0,0,+,-)  , (0,0,-,+) , (+,-,0,0) , (-,+,0,0) & 17.0949 \\


(0,+,0,+)  , (+,0,+,0) , (0,-,0,-)  , (-,0,-,0) & 17.0949 \\ \hline

(+,-,+,-) , (-,+,-,+) & 9.3997 \\ \hline\hline

(+,0,-,-) , (+,+,0,-)  , (-,0,+,+) , (-,-,0,+) & -0.1972 \\

(0,+,-,-) , (+,+,-,0)  , (0,-,+,+) , (-,-,+,0) & -0.1972
\end{tabular}

\label{coulombmatrix}
\end{table}
It is possible to use the same kind of symmetry arguments for the
calculation of the Coulomb matrix elements. Again we find that a
three-fold symmetry axis allows for more non-zero elements than a
$C_{\infty z}$-axis. Nevertheless, these additional matrix elements
are rather small. If the entire system is rotated by an angle
$\theta=\frac{2\pi}{3}$, each single-particle wave function acquires
a phase factor, while the distance $|\mathbf{r}-\mathbf{r}'|$ is not
affected by this rotation. Therefore we find for the interaction
matrix elements:
\begin{equation}
V_{ijkl}=e^{i(\lambda^z_i+\lambda^z_j-\lambda^z_k-\lambda^z_l)\theta}V_{ijkl}
\quad .
\end{equation}
From this one  can deduce that $V_{ijkl}$ must be zero if
$\lambda^z_i+\lambda^z_j-\lambda^z_k-\lambda^z_l \neq 0 \;
\text{modulo}\; 3$.
The $z$-projection of the angular momentum is conserved only modulo
3, and not exactly, as it would be in the case for a continuous
rotation axis $C_{\infty z}$.~\cite{BaGa04}. This gives rises to
additional non-zero matrix elements $V_{ijkl}$. Examples for the
electron-electron $V_{ijkl}$ are given in Tab.~\ref{coulombmatrix}
using here the ($s$,$p_+$,$p_-$) representation. Clearly these
additional matrix elements are rather small. This is in accordance
with our expectation about matrix elements that would altogether
vanish in a system with a higher symmetry.

Symmetry is also responsible for some Coulomb matrix elements having
equal values. Such cases are grouped in Tab.~\ref{coulombmatrix}
between horizontal lines.

\section{Many Body Problem}
\label{sec:manybodyprob}
%
%
\subsection{Hamiltonian}
So far we discussed only the single-particle properties. However,
the investigation of optical properties is an inherent many-particle
problem. The Hamiltonian $H$ that describes the system of
interacting electrons and holes in a QD consists of two parts:
\begin{align}
\label{eq:ehHamiltonian}
 H_0=&\sum_{i\sigma} \varepsilon^e_i e^{\dag}_{i \sigma}e_{i \sigma} +
      \sum_{i\sigma} \varepsilon^h_i h^{\dag}_{i \sigma}h_{i \sigma} \quad,
      \nonumber \\
 H_{\text{Coul}}=
      & \frac{1}{2} \sum_{ijkl \atop \sigma \sigma'} V^{ee}_{ij,kl} \;
      e^{\dag}_{i \sigma}  e^{\dag}_{j \sigma'} e_{k \sigma'} e_{l \sigma}
\nonumber\\
    &+ \frac{1}{2} \sum_{ijkl \atop \sigma \sigma'} V^{hh}_{ij,kl} \;
      h^{\dag}_{i \sigma}  h^{\dag}_{j \sigma'} h_{k \sigma'} h_{l \sigma}
\nonumber \\
    &-\sum_{ijkl \atop \sigma \sigma'} V^{he}_{ij,kl} \;
h^{\dag}_{i \sigma}  e^{\dag}_{j \sigma'} e_{k \sigma'} h_{l \sigma}
\quad.
\end{align}
The free part $H_0$ contains information about the single-particle
spectrum $\varepsilon_i^{(e,h)}$ and describes a system of
non-interacting charge carriers. Here $e_{i \sigma}$ ($e^{\dag}_{i
\sigma}$) are annihilation (creation) operators of electrons with
spin $\sigma$ in the one-particle  states $|i\rangle$ of energy
$\varepsilon^e_i$. The corresponding operators and single-particle
energies for holes are $h_{i \sigma}$ ($ h^{\dag}_{i \sigma}$) and
$\varepsilon^h_i$, respectively. The electron-electron, hole-hole
and electron-hole Coulomb interaction between the charged carriers
is included in $H_{\text{Coul}}$. The Coulomb matrix elements
$V_{ij,kl}$ are determined from the TB single particle wave
functions, as described in Sec.~\ref{interactionmatrix}.

%
%
\subsection{Configuration interaction (CI)}

\label{fulldiag}
In a semiconductor QD the finite height of the confinement potential leads to a
finite number of localized states as well as to a continuum of energetically
higher delocalized states. If one considers only the energetically lowest
single-particle states, the eigenvalue problem for a given number of
electrons and holes has a finite
(albeit large) dimension and can be solved without further approximations.

As the Hamiltonian $H$ conserves the total number of electrons $N_e$ and
holes $N_h$, the Hamiltonian matrix falls into subblocks with basis
states corresponding to uncorrelated many-particle states of the form
\begin{equation}
\label{eq:BasisEH} | \phi \rangle=\prod_{\substack{i}}
       (e^{\dag}_i)^{n^{e}_i}
               \prod_{\substack{j}}
       (h^{\dag}_j)^{n^h_j} \,\, |0\rangle \quad ,
\end{equation}
with the occupancy numbers $n^{e,h}_i=0,1$ and $\sum_i n_i^e=N_e$,
$\sum_j n_j^h=N_h$. In order to find the eigenvalues and
eigenfunctions of the interacting problem, the Hamiltonian $H$,
Eq.~(\ref{eq:ehHamiltonian}), is expressed in terms of the
uncorrelated basis states $|\phi\rangle$, Eq.~(\ref{eq:BasisEH}).
The resulting Hamiltonian matrix is then diagonalized numerically.
In this way one finds an expansion of the interacting eigenstates
for a given number of electrons and holes in terms of the
uncorrelated basis states of the system. These states can be used to
calculate, for example, the interband emission spectra between the
interacting Coulomb correlated eigenstates of the QD system using
Fermi's golden rule:
\begin{equation}
\label{eq:Fermi} I(\omega)={2\pi \over \hbar} \sum_f |\langle
\phi_f| H_{\text{d}} |\phi_i \rangle |^2 \,\,
\delta(E_i-E_f-\hbar\omega) \quad .
\end{equation}
Here $|\phi_i\rangle$  denotes the correlated initial state with
energy $E_i$ and $|\phi_f\rangle$ and  $E_f$ the corresponding
quantities of the final states. A similar equation holds for the
absorption spectrum. The Hamiltonian $H_{\text{d}}$ describes the
light matter interaction in dipole approximation
\begin{equation}
\label{eq:Hdipol} H_{\text{d}} = -e|\vec{E}|\sum_{n,m} \langle n|
\rvec \vec{e}_{p}|m\rangle h^{\dag}_{n,\sigma} e^{\dag}_{m,-\sigma}
+ \text{h.c.} {,}
\end{equation}
with $\vec{E}$ being the electric field at the position of the QD,
$\vec{e}_{p}$ is the polarization vector,  and $e$ is the elementary
charge. Furthermore the states $|n\rangle$ and $|m\rangle$ denote
hole and  electron single-particle states, respectively. Fermi's
golden rule, Eq.~(\ref{eq:Fermi}), implies that the optical field
always creates or destroys electron-hole pairs. Hence the initial
and final states differ by exactly one electron-hole pair. For the
following discussion it is convenient to express the dipole
Hamiltonian in terms of the interband polarization operator $\P$
according to $H_\text{d}=-e |\vec{E}|(\P+\P^\dagger)$.

If the analysis is restricted to $s$- and $p$-shells and the small
dipole matrix elements $d^{eh}_{p_-p_+}$ and $d^{eh}_{p_+p_-}$ are
neglected, one can split $\P$ into $\P_\text{low}$ and
$\P_\text{high}$ according to: \eqalN{}{
\P_{ \text{high} }  & = \sum_\sigma \Big(  d^{eh}_{ps}\;e_{p_+,\s} h_{s,-\s} +  d^{eh*}_{ps}\; e_{p_-,\s}h_{s,-\s}\Big)\,, \\
\P_{ \text{low} } & = \sum_\sigma   \Big(  d^{eh}_{sp}\; e_{s,\s}
h_{p_+,-\s} +  d^{eh*}_{sp}\; e_{s,\s} h_{p_-,-\s}\Big)\,. }{} This
is motivated by the fact that the single-particle energy separation
for the electrons is larger than for the holes.

%
%

\section{\label{nitride_excitonic_properties}Optical Properties}

To be able to distinguish between effects stemming from the symmetry
group of the QD and those additionally introduced by the internal
electric field, we will first present the results where the built-in
field is artificially switched off. In a second step, we will
discuss the influence of this field on the multi-exciton spectra.

\subsection{Excitonic and biexcitonic properties}

\begin{figure}[t]
\includegraphics[width=\columnwidth]{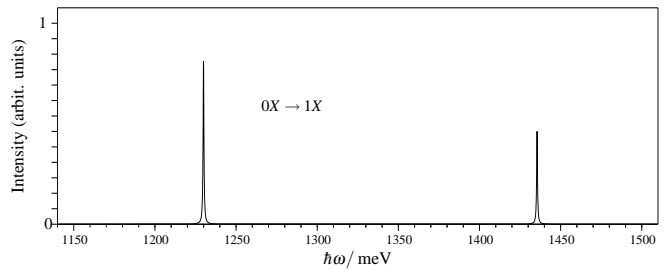}
\caption{Absorption spectrum of the initially empty nitride QD. The
high-energy side stems from a transition in which mainly the
$s$-shell of the hole and the $p$-shell for the electrons is
occupied. For the low-energy side the $s$-shell of the electrons and
the $p$-shell of the holes yield the main contribution.}
\label{NoPiezoAbs1S}
\end{figure}

The considered small QD confines only three bound electron states
($s$- and $p$-shell) for each spin polarization. These states enter
the CI approach, together with the hole $s$- and $p$-shells, which
are energetically well separated from the other localized states.
The corresponding excitonic absorption spectrum shows the two lines
depicted in Fig.~\ref{NoPiezoAbs1S}. Here we denote an electron-hole
pair, in which the electron has mainly $\a$ and the hole mainly $\b$
character, as an $\a\b$-exciton with $\a,\b\in\{s,p\}$. We found in
Section~\ref{sec:dipolesymm} that in the nitride case the
transitions do not originate from 'diagonal' $ss$- and
$pp$-excitons, as one may expect, but from $sp$- and $ps$-excitons.
Because of the large energy splitting between the $s$- and the
$p$-shell of the electrons, the $ps$-excitonic transition is well
separated from the $sp$ transition. Without Coulomb interaction,
these transitions can be found at the sum of the single-particle
energies $\ep_p^e+\ep_s^h$ and $\ep_s^e+\ep_p^h$, respectively. For
the absorption spectrum the unusual selection rules described in
Sec.~\ref{sec:dipolesymm} and the resulting ``skew'' excitons lead
only to quantitative changes. Even if the optical spectra could be
described in terms of diagonal excitons, one would still obtain two
lines in the absorption spectrum \cite{Haw99}, one from the
$s$-shell and one involving the $p$-shell carriers. However, for the
emission spectrum the changes are quite important. Most importantily
the exciton and biexciton ground state for small InN/GaN QDs remain
dark.~\cite{BaSc05} This is the case because the excitonic and
biexcitonic ground states are dominated by those configurations
where all the carriers are in their energetically lowest shell
together with the fact that the dipole-matrix element $d^{eh}_{ss}$
involving these states vanishes.

\subsection{\label{nitride_FCI}CI results for multi-exciton emission spectra}

\begin{figure}[t]
\begin{center}
\includegraphics{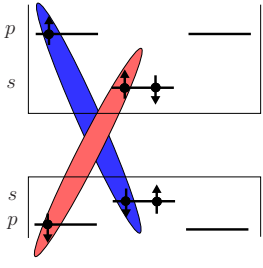}
\end{center}
\caption{\co For the example of a three exciton state, the light
(red) and dark (blue) shaded areas connect the carriers that will
lead to an emission at the low- and high-energy side of the
spectrum, respectively.} \label{reversedlevel}
\end{figure}

Only for more than two excitons the CI calculation provides
significant population of the electron and hole $p$-shells. Then
emission processes involving the \quot{skew} excitons can take
place. As mentioned above the low- and high-energy side of the
spectrum can be attributed to processes where an $s$-electron or a
$p$-electron is dominantly involved in the recombination process,
respectively. A schematic representation of the level-structure and
electron-hole pairs typically involved in high- and low-energy
transitions is depicted in Fig.~\ref{reversedlevel}.

An inspection of the emission-spectra reveals a  blue-shift as the
number of excitons is increased.  This is in strong contrast to the
results known from the InGaAs system \cite{BaSt00, DeGe00, BaGa04}
but can be explained in terms of the diagonal Hamiltonian, discussed
in the next section, and the fact that the envelopes for the
electrons and holes differ strongly. Also it can be seen that, with
the exception of the $5X\rightarrow4X$ transition, all  spectra are
rather similar if one compares the line structure of the low- and
high-energy side. On the other hand, the oscillator strengths of the
peaks on the high-energy side are systematically weaker than the
corresponding ones at the low-energy side. These aspects will be
addressed in the following in detail.

\subsection{\label{nitride_diag}Diagonal approximation}

As discussed in Ref.~\onlinecite{BaGa04}, an approximate description
using a Hamiltonian that is diagonal in the free states can be
motivated by inspecting the relative importance of the various
Coulomb matrix element. The eigenvalues of this approximate
Hamiltonian are then the non-selfconsistent Hartree-Fock energies.
This approximation has successfully been used to describe the
dominant trends in multi-exciton spectra in III-V
systems.~\cite{DeGe00, FrWi00, ShFr01, BeZu03, UlBe05}

\begin{figure}
\includegraphics[width=\columnwidth]{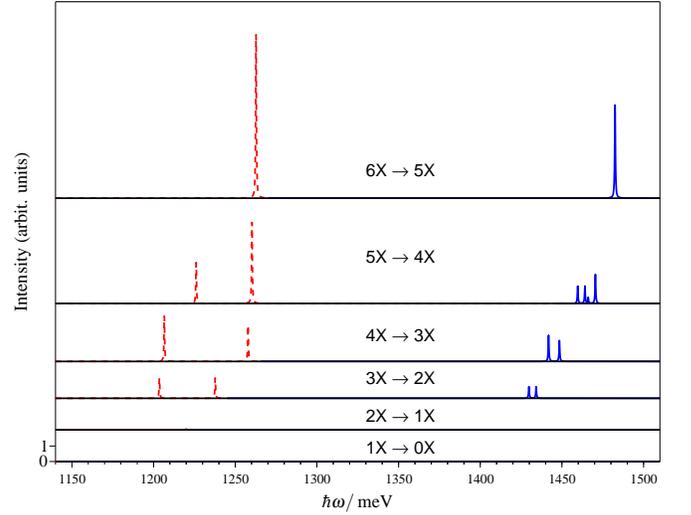}
\caption{\co Ground state emission spectra for a QD with different
number of excitons. The high-energy side is shown in blue (solid
lines), the low-energy side in red (dashed lines). For the studied
system almost no ground state emission is observed for exciton and
biexciton. As initial states the ground states with $S^{tot}_z=0$
are chosen and the internal electric field is switched off.}
\label{NitrideSmallQDFull_NoPiezo_br}
\end{figure}

\begin{figure}
\includegraphics[width=\columnwidth]{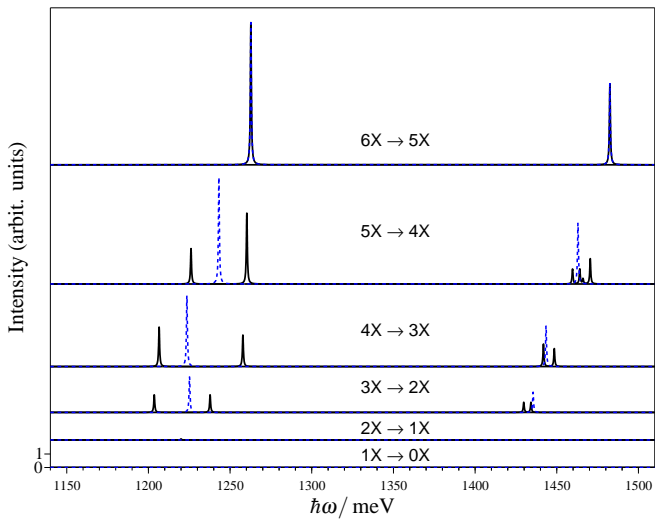}
\caption{(color online) Comparison of QD ground state emission
spectra using the CI-approach (solid lines) and  the diagonal
Hamiltonian (dashed lines) for various multi-exciton transitions. In
case of six excitons the two results practically coincide. All data
are calculated in the absence of the piezo electric field.}
\label{NitrideSmallQDFull_and_approx_NoPiezo}
\end{figure}

In a ground state emission spectrum, without Coulomb interaction,
one would  observe one line at the high-energy side at
$\epsilon^e_p+\epsilon^h_s$ and one at $\epsilon^e_s+\epsilon^h_p$
on the low-energy side  for an initial filling from three to six
excitons. With Coulomb interaction,  one observes instead of single
lines clusters of peaks as shown in
Fig.~\ref{NitrideSmallQDFull_NoPiezo_br}. These clusters blue-shift
as the number of excitons is increased. The approximate position of
the clusters can already be explained by considering only the
diagonal elements of the Hamiltonian matrix, see
Fig.~\ref{NitrideSmallQDFull_and_approx_NoPiezo}. Note that
non-identical envelopes lead to additional interaction terms in
comparison to Refs.~\onlinecite{Haw99,DeGe00,BaGa04}. The transition
energy $\Delta E^{\text{diag}}_{A\rightarrow B}$ from a
many-particle configuration $\ket{A}$ to a configuration $\ket{B}$
is given to a first approximation by~\cite{BaGa04}
\eqalN{Diag_nitride}{
\Delta E^{\text{diag}}_{A\rightarrow B} &= \epsilon^e_{\bar{e}} + \epsilon^h_{\bar{h}} - D^{eh}_{\bar{e}\bar{h}} \;\\
& +  \sideset{}{} \sum_{i\neq \bar{e},\bar{h}} \big( D^{ee}_{\bar{e} i} \, n_i^e - D^{eh}_{\bar{e} i } \, n_i^h - X^{ee}_{\bar{e} i } \, n_i^e \big) \\
& +  \sideset{}{} \sum_{i\neq \bar{e},\bar{h}} \big( D^{hh}_{i\bar{h} } \, n_i^h - D^{eh}_{i \bar{h} } \, n_i^e - X^{hh}_{i \bar{h} } \, n_i^h \big)
}{.}
Here $\bar{e}$ and $\bar{h}$ denote  the single-particle states of
the electrons and holes, respectively, that are depopulated in the
emission process. The index $i$ involves all single-particle states
except $\bar{e}$ and $\bar{h}$. The quantity
$D^{\lambda\lambda'}_{ij}$ stands for the direct Coulomb matrix
elements $V^{\lambda\lambda'}_{ijji}$, while the exchange matrix
elements $X^{\lambda\lambda'}_{ij}$ are given by
$V^{\lambda\lambda'}_{ijij}$. Of course, these exchange terms
contribute only if the spin of $\bar{e}$ or $\bar{h}$ agrees with
the electron or hole state $i$.  The first line of
Eq.~(\ref{Diag_nitride}) contains the free particle energies
$\epsilon^e_{\bar{e}}$ and $\epsilon^h_{\bar{h}}$ of the recombining
carriers together with the attractive Coulomb interaction matrix
element $- D^{eh}_{\bar{e}\bar{h}}$. All terms that stem from the
interaction of the electron in state $\bar{e}$ with all the other
electrons and holes have been grouped in the second line. Similarly,
the third line contains the interaction between the hole labeled
with $\bar{h}$ and all the other carriers. Explicitly, one obtains
for the high-energy transition of the $3X$ configuration:
\eqalN{}{ E^{\text{diag}}_{3X\rightarrow 2X} &= \epsilon^e_p +
\epsilon^h_s - D^{eh}_{ps} \\
&+ 2 D^{ee}_{sp}- D^{eh}_{ps}- D^{eh}_{pp}-X^{ee}_{sp}
\\
&+ D^{hh}_{ss}+ D^{hh}_{sp} - 2 D^{eh}_{ss}-X^{hh}_{sp} }{.}

Similarly, one finds from Eq.~(\ref{Diag_nitride}) for the ground state transitions on
the high-energy side for more than three excitons:
\eqalN{NitrideDiagonal}{
E^{\text{diag}}_{4X\rightarrow 3X} & = E^{\text{diag}}_{3X\rightarrow 2X} + \Delta
E_{\text{Hartree}} - X^{ee}_{pp} - X^{hh}_{sp}\,, \\
E^{\text{diag}}_{5X\rightarrow 4X} & = E^{\text{diag}}_{4X\rightarrow 3X} + \Delta
E_{\text{Hartree}}\,,  \\
E^{\text{diag}}_{6X\rightarrow 5X} & =
E^{\text{diag}}_{5X\rightarrow 4X} + \Delta E_{\text{Hartree}}\,.
}{} Note that the Hartree-shift $\Delta E_{\text{Hartree}}$, defined
by \mbox{$( D^{ee}_{pp} - D^{eh}_{pp} ) + ( D^{hh}_{ps} -
D^{eh}_{ps} ) $} , would be zero for identical envelopes of
electrons and holes. The matrix element $X^{ee}_{pp}$ is given by
$V^{ee}_{p_+p_-p_+p_-}$.  The peaks obtained from the diagonal
description provide the approximate position of the clusters
calculated by the CI, see
Fig.~\ref{NitrideSmallQDFull_and_approx_NoPiezo}. In particular the
smaller shift of the $4X$ spectrum relative to the $3X$ spectrum as
compared to the shifts involving more excitons is well described in
terms of the exchange matrix elements present in the first line of
Eq.~(\ref{NitrideDiagonal}).  The corresponding transition energies
of the low-energy side can be found by changing $e\leftrightarrow h$
in the equation above.

While the central position of the clusters is well reproduced by the diagonal
treatment, the different splitting within each cluster is not explained. A detailed
analysis of the states involved in the different emission processes shows that a
semi-analytic description of the spectrum is possible. Such a description will be given
in the following sections.


\section{Semi-Analytic Discussion of the Multi-Exciton Spectra}
\label{sec:semianalytic}

In the past, optical multi exciton spectra for nitride QDs have only
been discussed with selection rules carried over from the InGaAs
system.~\cite{HoFe99, Shi02, RiAm04} Additionally, the diagonal
treatment, which is often suitable for the description of major
trends in the InGaAs system~\cite{DeGe00, FrWi00, ShFr01, BeZu03,
UlBe05}, can in the present case reproduce only the overall position
of the clusters, but is by no means sufficient to explain the
multiplets. Therefore we will discuss in the following the different
multi-exciton spectra of \epsr{NitrideSmallQDFull_NoPiezo_br} in
detail.


\subsection{$\mathbf{3X\rightarrow 2X}$ Emission spectrum}
\begin{figure}[t]
\includegraphics[width=\columnwidth]{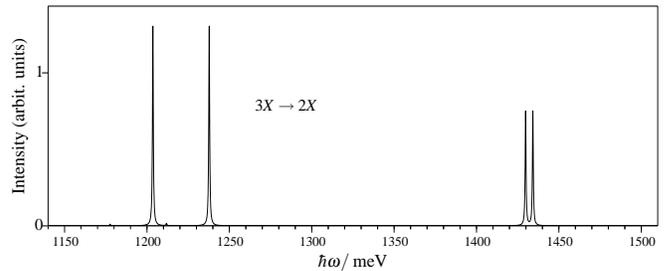}
\caption{\label{NoPiezo3} Ground state emission spectrum for an
initial filling of three excitons. As initial states the ground
states with $S^{tot}_z=0$ are chosen.}
\end{figure}

The ground state emission spectrum with an initial filling of three
excitons, is dominated by pairs of lines on the high- and on the
low-energy side, see Fig.~\ref{NoPiezo3}. We will analyze only the
low-energy side, as the same line of arguments can be carried out
for the high-energy side.

Due to the spin symmetry we can restrict ourselves to one of the two
possible ground states, for which the dominant configuration
(according to the CI calculation) is shown in \epsr{3XGs}. A linear
combination of configuration $\ket{A}$ and $\ket{B}$, enters the
$3X$-ground state with the same amplitude but opposite sign.
Nevertheless, one can restrict the discussion to either $\ket{A}$ or
$\ket{B}$, because they individually produce the same spectrum and
possible interference terms in the expansion of the coupling matrix
elements appearing in Fermi's golden rule are zero.

\begin{figure}[b]
\bc
\includegraphics[totalheight=2.5cm]{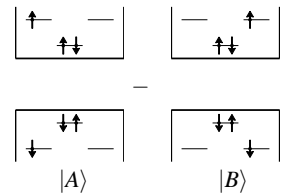}
\caption{\label{3XGs} One of the two configurations dominating the
three-exciton ground states with $S^{tot}_z=0$. } \ec
\end{figure}

In \epsr{Config3X} configuration $\ket{A}$ is shown together with the
$2X$-configuration (in black) that can be reached via the action of  $\P_{\text{low}}$.
Note  that the latter is not an eigenstate of the Hamiltonian. Instead it will be mixed
with other states by Coulomb interaction. The predominant contribution to this mixture
is shown in \epsr{Config3X} in light grey. Together these two states either form a
spin triplet state for the electrons and a spin singlet for the holes ($ts$) or a
singlet-singlet ($ss$)  state. These $ss$ and $ts$ states are split by the exchange Coulomb
matrix element $2X_{sp}^{ee}$ and can both be observed in the spectrum. The oscillator
strengths of the corresponding transitions are equal since both final
states contain the bright $2X$-exciton state with the same probability amplitude. Along
the same line one finds for the high-energy side an approximate splitting of
$2X_{sp}^{hh}$ which is, however, considerably smaller. Both splittings are in good agreement with the CI result and explain
the dominant peak structure in \epsr{NoPiezo3}. The ratio of the
peak heights on the low- and the high-energy side is given in terms of the dipole-matrix elements by
$d^{eh}_{sp}/d^{eh}_{ps}$.

\begin{figure}[t]
\begin{center}
\includegraphics[totalheight=3.15cm]{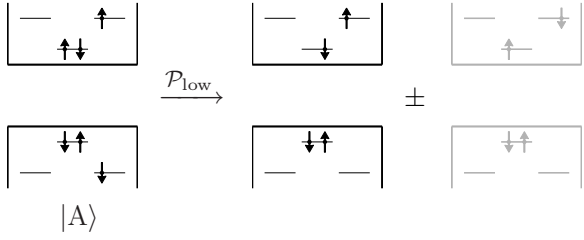}
\caption{\label{Config3X} One part of the dominant configuration of
the $3X$-ground state characterized by $S_z^e=+\half$ and
$S_z^h=-\half$, together with the $2X$ configuration (in black) that
is created by removing the $sp$-exciton via the action of
$\P_{\text{low}}$. Additionally the most important dark
configuration to which the bright $2X$-state couples via Coulomb
interaction is shown in light grey.}
\end{center}
\end{figure}
%


\subsection{$\mathbf{4X\rightarrow 3X}$ Emission spectrum}
The ground state emission spectrum of $4X$ looks similar to the $3X$
emission spectrum. However, the energetic splitting within the
cluster is larger and the oscillator strength of the two lines of
the cluster differs from each other. By considering the involved
states, we find that the difference can be explained by the fact
that the final states are now doublet-doublet ($dd$) and
quadruplet-doublet ($qd$) states and no longer simple $ss$ and $ts$
states as in the $3X$ case. The corresponding quadruplet-doublet
splitting, which determines the energy difference within the
cluster, is larger than the singlet-triplet splitting. Evaluating
the contribution of the three different $S_z^\text{tot}=0$ states,
yields an approximate ratio of 5 to 4 for the different oscillator
strengths within the cluster and is in good agreement with the CI
result.


\subsection{$\mathbf{5X\rightarrow 4X}$ Emission spectrum}


\begin{figure}[b]
\includegraphics[width=\columnwidth]{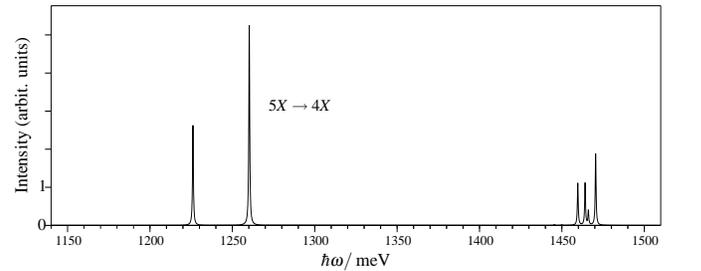}
\caption{Ground state emission spectrum for an initial filling of
five excitons. As initial states the ground states with
$S^{tot}_z=0$ are chosen.} \label{NoPiezo5}
\end{figure}

In contrast to the other multi-exciton spectra, Fig.~\ref{NoPiezo5}
shows a clear asymmetry and additional lines at the high-energy side
of the spectrum.
\begin{figure}[!t]
\bc
\includegraphics[totalheight=2.5cm]{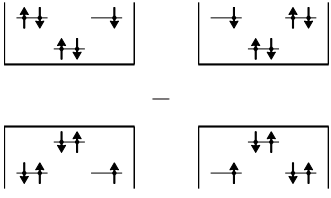}
\caption{\label{5XGs} Dominant ground state configuration for five
excitons with $S^{tot}_z=0$. By flipping all spins a second,
degenerate ground state is produced.} \ec
\end{figure}
One of the $5X$ ground states is depicted in \epsr{5XGs}. Taking
this as the initial state,  the main contribution to the final
states for the high-energy side stem from the $6 \times 6$ block
generated by the configurations schematically represented in
\epsr{Config4XFinal}. These states have the lowest non-interacting
energy amongst those states that can be reached by a removal of one
$ps$-exciton from the configurations shown in \epsr{5XGs}. Their
electronic configuration is given by $2 e_s 2 e_p 1h_s 3h_p$. A
similar block with configurations $1 e_s 3 e_p 2h_s 2h_p$ is found
for the low-energy transitions. Due to this symmetry one would
expect the same number of lines on the high- and on the low-energy
side of the spectrum. This expectation is fulfilled for all but the
$5X\rightarrow 4X$ transitions.

%
\begin{figure}[b]
\bc
\includegraphics[totalheight=2.95cm]{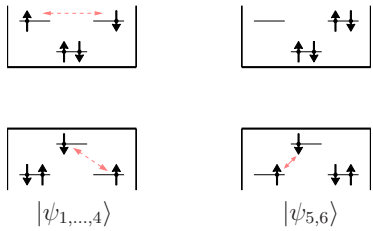}
\caption{\label{Config4XFinal} \co Main contribution to the final
states of the high-energy side of the $5X\rightarrow4X$ transition
with classification $S^\text{tot}_z=0$ and, assuming full angular
momentum conservation, $l^\text{tot}_z=0$. The arrow between
electrons or holes indicates that additional states can be derived
from the displayed configuration by flipping the spins of the
connected carriers simultaneously. This way one can create four
different states from the first configuration and two from the
second one. } \ec
\end{figure}


By combining the first four states $\ket{\psi_{1,\ldots,4}}$ in
\epsr{Config4XFinal} one can form  ss, st, ts and tt spin states.
The last two states $\ket{\psi_{5,6}}$ in \epsr{Config4XFinal} allow
the  formation of an st and ss state. While the electrons of the
first four states always occupy both $p_+$ and $p_-$, they occupy
only the $p_-$ state in the last two configurations. Therefore one
expects that the energy of the $ss$ state formed  by the states
$\psi_{1}$ to $\psi_{4}$ differs from the energy of the eigenstate
created by combining $\psi_{5}$ and $\psi_{6}$. By the same token
one expects two different energies for the two possible $st$-states.
Therefore six different energies can be expected. On the high energy
side of the spectrum four lines are clearly visible and another two
may be identified on the left side of the cluster. For the
low-energy side, however, only two lines can be observed.

To obtain further insight,  we construct the Hamiltonian matrix
generated from the states $\{|\psi_i\rangle\}^6_{i=1}$ and
block-diagonalize it by transforming to spin-eigenstates. This way
one obtains four subblocks: \eqa{}{
H_\text{ts} & = -\frac{3}{2} \;, & \qquad H_\text{{ss}}  = \half\begin{pmatrix}-1 & t\\t& 1 \end{pmatrix}\;, \\ 
H_\text{st} & = -2\tilde{t}\uu{1} - H_{ss}\;,  &  \qquad H_\text{tt}     = -2\tilde{t} -\frac{3}{2} \;,
}{}
where we introduced the dimensionless
parameters $t=2\sqrt{2}X^{eh}_{pp}/X^{ee}_{pp}$ and
$\tilde{t}=X^{eh}_{pp}/X^{ee}_{pp}$ and measured all energies in units of
$X^{ee}_{pp}$ relative to $E^\text{diag}+\tilde{t}+\half$ with $E^\text{diag}$ being the energy of the configurations in the diagonal approximation. The six eigenvalues of these blocks are $E_\text{ts} = -\frac{3}{2}$, $ E_\text{ss}=\pm \half \sqrt{1+t^2}$,  $E_\text{st} = -2\tilde{t} \mp \half \sqrt{1+t^2}$, and
$E_\text{tt}=-2\tilde{t} -\frac{3}{2}$.  From these expression one can read off that the
st and tt spectrum is shifted by $-2\tilde{t}$ relative to the $ss$ and $ts$
spectrum, respectively.

\begin{figure}
\begin{center}
\includegraphics[totalheight=5.1cm]{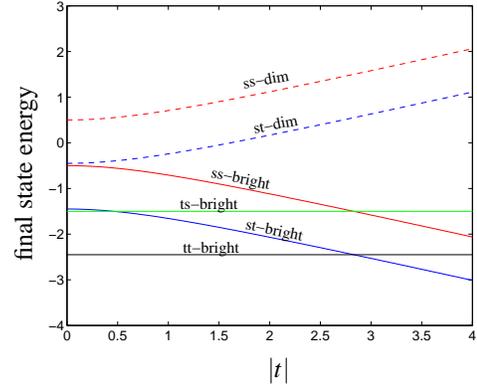}
\end{center}
\caption{\label{eigenenergies} \co Energies of the final states
involved in the high-energy transition measured in units of
$X^{ee}_{pp}$ relative to $E^\text{diag}+\tilde{t}+\half$ as a
function of the dimensionless parameter $t  = 2\sqrt{2}
{X^{eh}_{pp}}/{X^{ee}_{pp} }$ for fixed $\tilde{t}\approx0.47$. The
labeling refers to the spin configuration of the final states and
whether the states are 'bright' or 'dim'. }
\end{figure}

In order to obtain the corresponding oscillator strengths of the
transitions, one has to calculate the matrix elements $|\bra{4X,i}
\P_{\text{high}} \ket{5X,\text{gs}}|^2$, where $\ket{4X,i}$ denotes
the $i$-th eigenstate of the four exciton problem and
$\ket{5X,\text{gs}}$ refers to the ground states schematically shown
in \epsr{5XGs} of the five exciton problem. Because the eigenstate
within the one-dimenstional subspaces of the ts- and tt-states is
not affected by a varied parameter $t$, the  oscillator strength
does not depend on the value of $t$. In contrast, the heights of the
ss- and st-lines depend strongly on $t$ as the amplitude of the
different states in the linear combination varies with $t$. Denoting
the eigenstates of $H_\text{ss}$ with $(\alpha_i,\beta_i)^t$  the
oscillator strength $\Osc$ of the ss transition is given by \eq{}{
\Osc=|d^{eh}_{ps}|^2 \; \Big|\frac{\alpha_i}{\sqrt2} +
\frac{\beta_i}{2} \Big| ^2 }{.}

\begin{figure}[b]
\begin{center}
\includegraphics[totalheight=4.9cm]{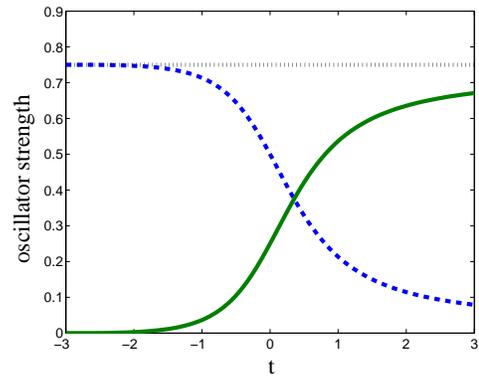}
\end{center}
\caption{\label{Oscillatorstrength} \co Oscillator strengths, shown
in green (solid line) and blue (dashed line), of the two  $ss$
transitions as a function of the dimensionless parameter $t$.  A
clear redistribution of oscillator strength from one to the other
transition is visible. The total oscillator strength, red (dotted
line), however, remains unchanged. The same behavior is found for
the st transition. }
\end{figure}

A similar analysis can be performed for the $st$ transition. In this case one finds
\eq{}{
\Osc=|d^{eh}_{ps}|^2 \; \Big|\frac{\alpha_i}{2} - \frac{\beta_i}{\sqrt2} \Big| ^2
}{}
for the oscillator strength. As $H_\text{ss}$ and $H_\text{st}$ have the same eigenvectors and if
$v=(\alpha,\beta)^t$ is an eigenvector, so is $v'=(-\beta,\alpha)^t$, one finds for the
$st$-transitions the same dependency of the oscillator strength as for the $ss$ transition.
For the analysis of the low-energy cluster, only  the labels $e\leftrightarrow h$ have to be changed
in all the derived equations.

The dependence of the transition energies and the oscillator
strengths on the parameter $t$ are depicted in \epsr{eigenenergies}
and \epsr{Oscillatorstrength}, respectively. For above discussed QD
example, one obtains $t\approx-1.2$ for the high-energy side and
$t\approx-2.8$ for the low energy side. Therefore we expect from
\epsr{Oscillatorstrength} that one will only be able to observe in
the spectrum one of the $ss$ and one of the $st$ lines in addition
to the $ts$ and $tt$ line. This explains the four clearly visible
lines in \epsr{NoPiezo5} of the high energy side. In contrast, for
the low-energy side only two peaks are visible. This can be traced
back to the fact that one has in this case $X^{eh}_{pp}  \approx
X^{ee}_{pp}$ or $t\approx 2\sqrt{2}$. According to
\epsr{eigenenergies} this means that the $ss$ and $st$ as well as
$ts$ and $tt$ have almost identical transition energies. As a
consequence of this degeneracy only two distinct lines can be
observed on the low-energy side of the spectrum.


\subsection{$\mathbf{6X\rightarrow 5X}$ Emission spectrum}

\begin{figure}
\includegraphics[width=\columnwidth]{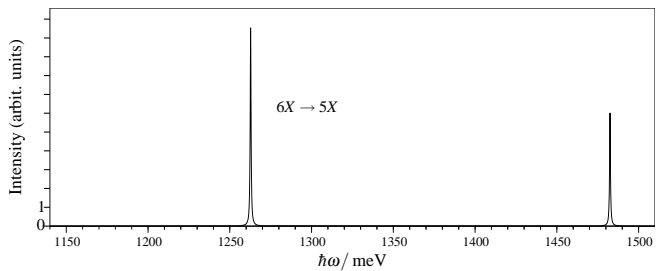}
\caption{Ground state emission spectrum for an initial filling of
six excitons.} \label{fig:6X}
\end{figure}

In the case of two shells for each type of carrier, the QD is completely filled with
six excitons and there is
only one ground state possible. By removing  one $sp$-exciton or one $ps$-exciton from
this configuration, we find
that the possible final states are exactly those for the $0X\rightarrow 1X$ transitions, only that
the occupied sites in the $0X\rightarrow 1X$  problem are now the \emph{un}occupied
ones. Therefore the  $6X\rightarrow 5X$ emission spectrum, see Fig.~\ref{fig:6X}, is very similar to
the $0X\rightarrow 1X$ absorption spectrum, with the main difference being that the
lines are shifted due to the
interaction with the 'background' carriers.

\section{\label{field_vs_nofield}Influence of the Built-In Field}

Nitride QDs grown along the $c$-axis are characterized by the
presence of a strong internal electrostatic field which has a
component stemming from the spontaneous polarization and a part
generated by strain. When this field is included in the calculation
of the single-particle properties, the electron and hole wave
functions are spatially separated from each other which leads to a
reduction of the oscillator strength.~\cite{AnRe01, RiAm02,ScSc06}
Furthermore the single-particle gap and the Coulomb matrix elements
are altered. The resulting multi-exciton spectra with and without
the inclusion of the built-in field are compared in
\epsr{NitrideSmallQDFull_NoPiezo_Piezo}.
\begin{figure}[b]
\includegraphics[width=\columnwidth]{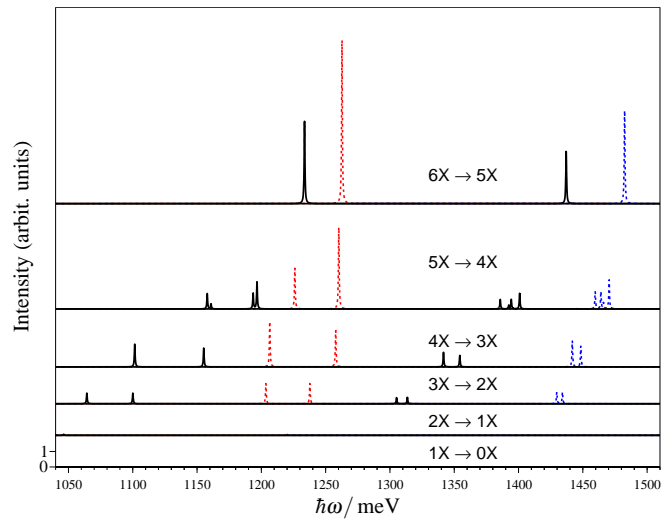}
\caption{\co Ground state emission spectra for the smaller QD ($d =
4.5$ nm and $h = 1.6$ nm)with (solid lines) and without (dashed
lines) the inclusion of the internal field. Different number of
excitons with $S^{tot}_z=0$ are chosen as initial states. For the
studied system almost no ground state emission is observed for the
exciton and biexciton.} \label{NitrideSmallQDFull_NoPiezo_Piezo}
\end{figure}
The reduced oscillator strength and the overall red-shift of the
spectra due to the QCSE are clearly visible. In addition to these
results of the modified single-particle properties, we find a
stronger shift of the lines with increasing number of excitons in
the presence of the internal field. This is another result of the
strong separation of the electron and hole wave functions which also
introduces Hartree-shifts in the spectrum. Indeed, if the diagonal
approximation is applied as outlined in Section \ref{nitride_diag},
one obtains again the position of the cluster to a good
approximation.  Another difference can be observed in the
$5X\rightarrow 4X$ spectrum: The number lines changes in the
presence of the intrinsic field. Since for all other excitonic
populations the spectra are only altered quantitatively this is a
rather surprising result. But this is resolved by noting that the
situation without the intrinsic field is rather special as one had
$X^{eh}_{pp} \approx X^{ee}_{pp}$. Including the built-in field
leads to a significant deviation of the two matrix elements and
henceforth  to a clear splitting of the previously degenerate lines.


\section{Multi-Exciton Emission for a Larger Quantum Dot\label{larger_dot}}

\begin{figure}[t]
\begin{center}
\includegraphics{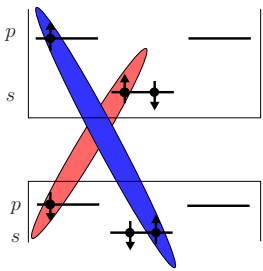}
\end{center}
\caption{ \co Schematic representation of a three exciton
configuration. The light (red) and dark (blue) shaded areas connect
the carriers that will lead to a emission at the low- and
high-energy side of the spectrum, respectively.} \label{scetchrev3x}
\end{figure}

In order to give a more representative overview, we additionally
investigate a larger QD with diameter $d=5.7\nm$ and height
$h=2.3\nm$. It turns out, that in this case the energetic order of
the two energetically lowest hole levels is reversed in the presence
of the internal electrostatic field. A schematic picture of this
situation is shown in Fig.~\ref{scetchrev3x}. In the absence of the
built-in field, one still has the \quot{usual} order with the
$s$-shell being lower in energy than the $p$-shell. Hence, the
spectrum looks similar to those previously discussed and is here
therefore omitted. However, in the presence of the built-in field
the two-fold degenerate $p$-shell constitutes the hole ground states
and has, according to the symmetry considerations of
Section~\ref{sym_wurtzite}, non-vanishing dipole matrix elements
with the electron ground state. This is in agreement with recent
$\vec{k}\cdot \vec{p}$ calculations \cite{Bagga2005} and
experimental results for CdSe QDs \cite{NiNo95} grown in the
wurtzite phase. As an immediate consequence the excitonic and
biexcitonic ground state is bright. However, the corresponding
$d^{eh}_{sp}$ dipole-matrix elements are strongly reduced in
comparison with the smaller QD due to the stronger separation of the
electron and hole wave function in this larger structure.

%
\begin{figure}[b]
\includegraphics[width=\columnwidth]{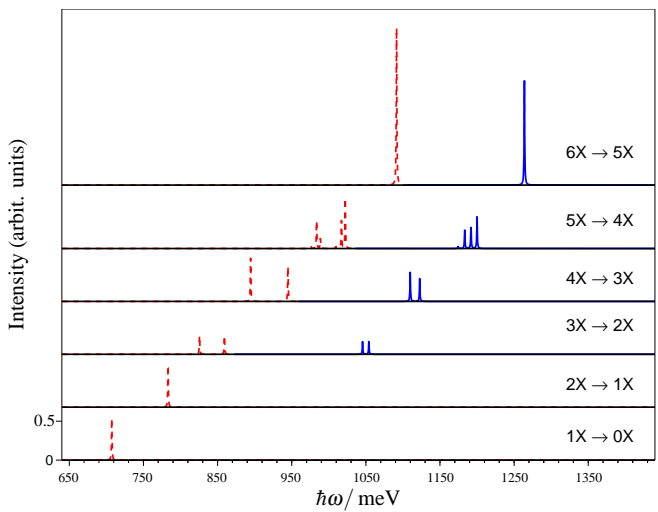}
\caption{(color online) Emission spectra for a larger quantum dot (d
= 5.7 nm and h = 2.3 nm) with different number of excitons and in
the presence of the internal field. For the emission of each
multi-exciton  complex the high-energy side is shown in blue (solid
lines), the low-energy side in red (dashed lines). Due to the
reversed level structure for the holes a ground state emission is
observed for exciton and biexciton. As initial states the ground
states with $S^\text{tot}_z=0$ are chosen.}
\label{NitrideLargeQDFull_Piezo}
\end{figure}

The resulting multi-exciton spectra are shown in
Fig.~\ref{NitrideLargeQDFull_Piezo}. In contrast to the intuitive
picture in which first the states with lowest single-particle energy
are occupied,  a strong population of the hole $s$-shell for more
than one exciton is found. The ground states are therefore not given
by those states with lowest non-interacting energy. This is already
confirmed by a calculation that contains only the Hartree Coulomb
terms and can qualitatively be explained as follows: The attraction
between the electron and hole being in their respective $s$-shells
is stronger than the attraction in the case of $s$- and $p$-shell
carriers. Therefore it can compensate the higher single-particle
energy of the hole in the $s$-shell. This leads already for the
biexciton to an occupation of the hole $s$-shell. However, an
additional promotion of the other hole is not favored in this case
because the increase in energy due to the stronger repulsive
interaction between the holes in their $s$-shell is higher than the
energy reduction due to the stronger attraction between
$s$-electrons and $s$-holes. As a consequence, the $s$- and
$p$-shell for the holes are equally populated in the biexciton
ground state. Therefore the biexcitonic line has approximately the
same oscillator strength as the excitonic one.

From three excitons on, the ground states are dominated by
configurations in which both the $s$-shell for electrons and the
$s$-shell for holes are fully populated. As a consequence, one
obtains qualitatively the same spectra as in the case of the smaller
dot with the \quot{normal} order of the shells. However, one finds
that the oscillator strengths are strongly reduced in the larger
system and that the Hartree shifts are even more pronounced, due to
the strong separation of electron and hole wave functions.

\section{Conclusion}
In this work we have investigated the electronic and optical
properties of lens-shaped InN/GaN QDs. Employing a tight-binding
model where the weak crystal-field splitting and spin-orbit coupling
for the system studied here are neglected, we found an exactly
degenerate single-particle $p$-shell. This degeneracy originates
from the $C_{3v}$-symmetry of the underlying wurtzite lattice. This
result is in particular intriguing in view of recent discussions in
the literature about the $p$-shell degeneracy in zinc-blende QDs.
Based on the microscopically determined single-particle wave
functions, the dipole and Coulomb matrix elements were evaluated.
These matrix elements served as input parameters for configuration
interaction calculations and allowed the determination and further
analysis of optical properties. Our prediction of dark exciton and
biexciton ground state for small dots is confirmed by symmetry
considerations. In contrast to other III-V material systems, the
emission from nitride-based QDs is dominated by 'skew' excitons, so
that completely different multi-exciton spectra arise. For larger
QDs, we found that the strong internal electric built-in field can
reverse the energetic order of the hole states, which results in a
bright exciton and biexciton ground state. However, the oscillator
strength is strongly reduced in these structures due to the quantum
confined Stark effect. By restricting the analysis to the
energetically lowest shells, a semi-analytic description of the
optical properties was possible, leading to a deeper insight into
the origin of the various emission lines.

\begin{acknowledgments}
We acknowledge financial support by the Deutsche
Forschungsgemeinschaft (research group ''Physics of nitride-based,
nanostructured, light-emitting devices``) and a grant for CPU time
from the NIC at the Forschungszentrum J\"ulich.
\end{acknowledgments}

\begin{appendix}

\end{appendix}

\bibliography{phd}

\end{document}